\documentclass[conference]{IEEEtran}
\IEEEoverridecommandlockouts

\ifCLASSINFOpdf
\else
\fi
\usepackage{tikz}
\usepackage{graphicx}
\usepackage{epstopdf}
\usepackage{float}
\usepackage{color}
\usepackage{amsmath}
\usepackage{array}
\usepackage{relsize}
\usepackage{subfigure}
\usepackage{amssymb,latexsym}
\usepackage{eso-pic}

\AddToShipoutPicture*{\small \sffamily\raisebox{1.2cm}{\hspace{1.8cm}To be published in Proc. of IEEE GLOBECOM 2016 First International Workshop on the Internet of Everything (IoE) \copyright 2016 IEEE}}
 
\begin{document}
%
\title{ Network Topology Modulation for \\  Energy and Data Transmission in \\ Internet of Magneto-Inductive Things}

\author{\IEEEauthorblockN{Burhan Gulbahar}
\IEEEauthorblockA{Applied Research Center of Technology Products\\
Department of Electrical and Electronics Engineering \\
Ozyegin University, Istanbul, 34794, Turkey\\
Email: burhan.gulbahar@ozyegin.edu.tr} 
\thanks{Dr. Burhan Gulbahar is with the Department of Electrical and Electronics Engineering and Applied Research Center of Technology Products, Ozyegin University, Istanbul, 34794, Turkey, (e-mail: burhan.gulbahar@ozyegin.edu.tr).   }
}

\maketitle

\makeatletter
\def\ps@headings{%
\def\@oddhead{\mbox{}\scriptsize\rightmark \hfil \thepage}%
\def\@evenhead{\scriptsize\thepage \hfil \leftmark\mbox{}}%
\def\@oddfoot{}%
\def\@evenfoot{}}
\makeatother
\pagestyle{empty}

\begin{abstract}
 
Internet-of-things (IoT) architectures connecting a massive number of heterogeneous devices need energy efficient, low {\color{black} hardware} complexity, low cost, {\color{black} simple} and secure mechanisms to realize communication among devices. {\color{black} One of the emerging schemes is to realize simultaneous wireless information and power transfer (SWIPT) in an energy harvesting network}.  Radio frequency (RF) solutions require special hardware and modulation methods for RF to direct current (DC) conversion and optimized operation to achieve SWIPT which are currently in an immature phase. {\color{black} On the other hand,}  magneto-inductive (MI) communication  transceivers are intrinsically energy harvesting with potential for SWIPT in an efficient manner. In this article, {\color{black}novel modulation and  demodulation mechanisms are presented in a combined framework with multiple-access channel (MAC) communication and wireless power transmission}. The network topology of power transmitting active coils in a transceiver composed of a grid of coils is changed as {\color{black} a novel} method to transmit information. {\color{black} Practical demodulation schemes are} formulated and numerically simulated for two-user MAC topology of small size coils. The transceivers are suitable to attach to everyday objects to realize reliable local area  network (LAN) communication performances with tens of meters communication ranges. The designed scheme is promising for future IoT applications requiring SWIPT with energy efficient, low cost, low power and low {\color{black}hardware} complexity solutions.
\end{abstract}




\section{Introduction}
\label{introduction}
 
{\color{black}The Internet of Things (IoT) and 5G architectures require connecting heterogeneous devices including machine-to-machine (M2M) and wireless sensor networking (WSN) units with potentially lower data rate but low latency, reliable, energy efficient and secure mechanisms \cite{related0}.} These applications require short packets and {\color{black}simple} modulation/demodulation mechanisms for the widespread utilization of IoT. The massive number of devices require not only low {\color{black} hardware complexity schemes but also methods of energy harvesting such as simultaneous wireless information and power transfer (SWIPT) transmitting data and energy by using radio frequency (RF) or magneto-inductive (MI)  methods \cite{related4, burhan3}}. However, the existing literature generally concentrates on the energy versus capacity trade-off including power allocation schemes \cite{related4, related2, related3}. The recent studies analyze RF based SWIPT modulation methods including spatial domain (SD) and intensity based energy pattern changes \cite{related4, related5}. RF solutions are not mature today to achieve SWIPT and require specialized circuits for RF to direct current (DC) conversion. In this article,  novel and practical network topology modulation and  demodulation architectures are presented for MI communications (MIC) networks by changing the spatial pattern of power transmitting active coils. The proposed MIC based IoT architecture (MI-IoT) provides reliable, {\color{black}simple} and secure mechanisms with low cost and low latency  performances for connecting everyday objects with direct power transmission capability.  

{\color{black}MIC is an alternative method with the  advantage of uniformity for varying environments without medium specific attenuation,  multi-path and high propagation delay in challenging environments including underwater, underground and nanoscale medium with in-body and on-chip applications \cite{ref1, ref2, ref4, ref5}.}  In \cite{related2} and \cite{related1}, a trade-off analysis is presented for the problem of information and power transfer on a coupled-inductor circuit with power allocation policies.  In \cite{ref5}, a nanoscale communication architecture with graphene coils is presented satisfying both power and data transmissions for in-body and on-chip applications. {\color{black}On the other hand, existing studies on  MIC networks  treat other coils as sources of interference including  multiple-input multiple-output (MIMO) and diversity architectures \cite{burhan3, ref1, ref2,  ref4, ref10, ref12}.}  However, SWIPT architectures,  MAC schemes utilizing the same time-frequency resources, and modulation methods other than classical signal waveform approaches are not discussed.
   
In this article, the information is embedded to coil positions by \textit{modulating the frequency selective MI channel} or \textit{network topology} {\color{black} instead of classical signal waveform modulation.} The proposed scheme fulfills the idea of fully {\color{black} coupled} information and power transfer for widespread utilization of MI-IoT applications \cite{related3}. It  eliminates {\color{black} signal modulation} and does not waste any power for data transmission.  {\color{black} Furthermore, it does not require transmitter pre-coding, channel state information and intensity pattern modulation as in RF counterpart utilizing Generalised Pre-coding aided Spatial Modulation (GPSM) scheme \cite{related5}.} Moreover, MAC sharing problem is intrinsically solved due to including whole network topology  as data symbol. {\color{black}The solution requires no receiver feedback and provides MAC sharing without sacrificing resources for transmitter synchronization or interference prevention.} The contributions achieved, for the first time,  are summarized as follows:
\begin{itemize}
\item {\color{black} Network topology modulation mechanism directly modulating the frequency selective MI channel.}
\item {\color{black} Practical topology demodulation combined with SWIPT.}
\item {\color{black}Topology modulating MACs fully utilizing time-frequency bands without synchronization or contention.}
\item Reliable, {\color{black}capable of} energy harvesting, low cost, low {\color{black} hardware} complexity  and low latency IoT  {\color{black}  networking}.
\end{itemize}
The proposed method supports the widespread adoption of MIC networks for IoT satisfying the following requirements:
\begin{enumerate} 
\item \textit{Low latency}: continuous energy and data transmission in a MAC topology without the overhead of resource sharing, synchronization  or receiver feedback.
\item \textit{High reliability}: the robustness of the MIC channel to fading and low probability of symbol error.
 \item  {\color{black} \textit{Low hardware complexity}: simple and low-cost SWIPT transceiver  with already available MI circuits, and without separate structures for data and power transmission.}
 \item  {\color{black} \textit{Energy harvesting capability}:} intrinsically by MI coils without requiring separate RF to DC conversion circuits.
 \item {\color{black} \textit{Security}: immunity to RF fields and radiative effects; potential detection of intruder coils  as the changes in network topology symbol and power transmission levels.}
\item \textit{Energy Efficiency}: {\color{black} no extra energy for signal waveform based data transmission;}  the crowded set of MI coils forming a waveguide to enhance the communication range without consuming any active transmission power.
 \end{enumerate} 
 
The remainder of the paper is organized as follows. In Section \ref{sysmodel}, MI network system model for MAC topologies is presented. In Sections \ref{stm} and \ref{stdm}, topology modulation and demodulation mechanisms are introduced, respectively. Then, in Section \ref{numersim}, the proposed methods are simulated for two-user MAC network.  Finally, in Sections \ref{openis} and \ref{conclusion}, open issues and conclusions are discussed, respectively.

\section{System Model}
\label{sysmodel}

\begin{figure}[!t]
\centering
\includegraphics[width=1.65in]{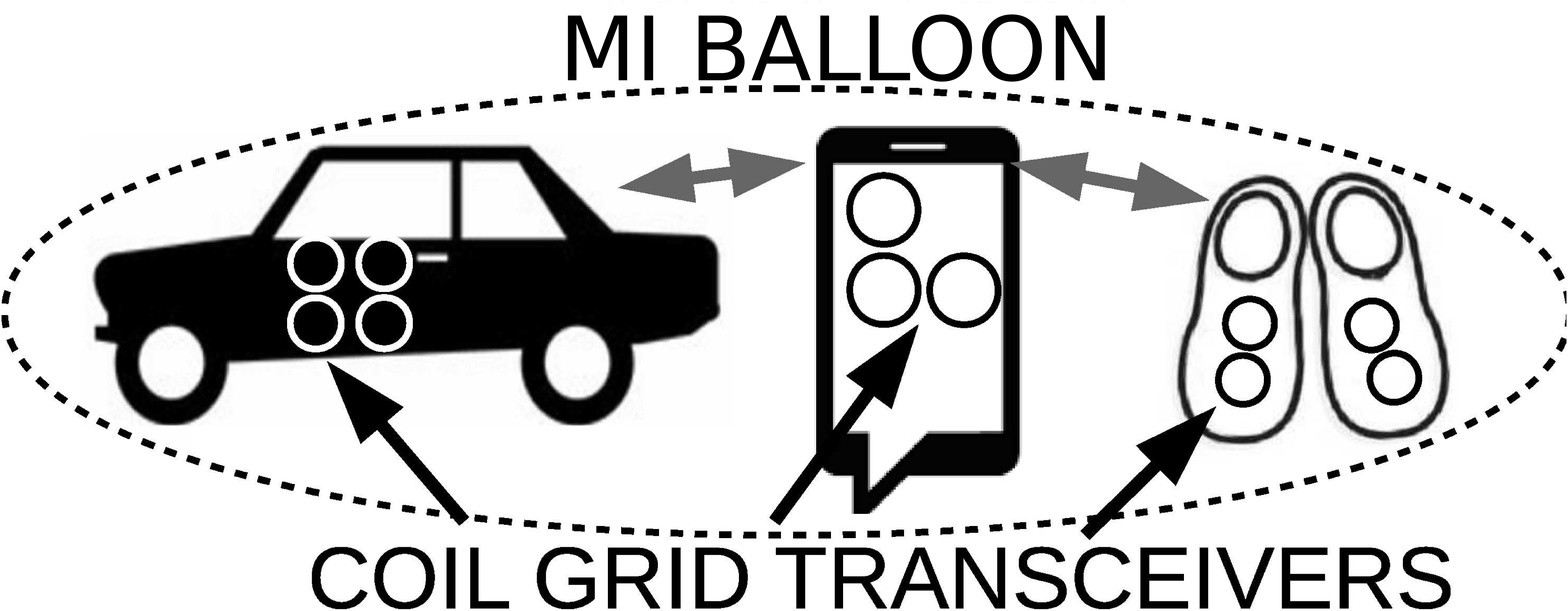} \hspace{0.1in} \includegraphics[width=1.65in]{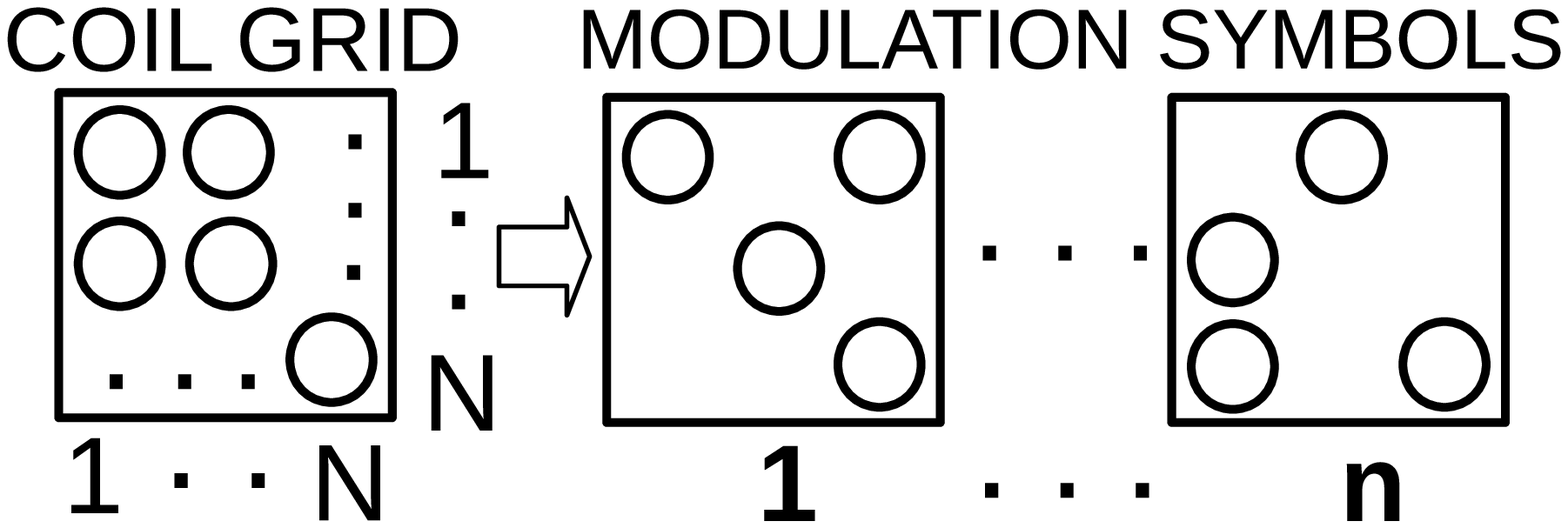} \\
(a) \hspace{1.4in} (b) \\
\caption{ (a)  {\color{black} MI-IOT for LANs of daily} objects forming a  coupled MI balloon, and (b) {\color{black} a grid of $N \times N$  coils at each transceiver with varying topologies consisting of $n$ different spatial symbols.}}
\label{fig12}
\end{figure}

In this paper, there are $K \, - \, 1$ transmitters denoted by $Tx_{j}$ for $j \in [1, \, K\, - \,1]$ to transmit data to a single receiver, i.e., indexed as the first user and denoted by $Rx$, in a MAC topology {\color{black} with  random positions and orientations as shown in Fig. \ref{fig12}(a).}  Each user {\color{black} has a} grid of closely spaced $N \times N$ coils where each coil is either in open circuit condition without any effect to the network or actively resonating as shown in Fig. \ref{fig12}(b). {\color{black} The number of concurrently active coils in each transceiver with the ability to change its spatial coil distribution is denoted by $T$. The grid can be realized on a flexible  paper suitable to attach to daily objects and to be recycled considering the billions of devices.}

{\color{black}Circuit theoretical equivalents of transceivers are presented in detail in \cite{burhan3, ref1, ref5}.  Each coil has series RLC type circuit with resistance $R_i$, capacitance $C_i$ and inductance $L_i$ for $i \in [1, \, N_{Tot}]$  where $N_{Tot} = K \times T$ with resonance frequency $\omega_0 = 1 \, / \, \sqrt{L_i \, C_i}$. Furthermore, each coil has the same properties  to simplify the analysis, i.e., $R_i = R$, $L_i = L$ and $C_i = C_s$ for $i \in [1, \, N_{Tot}]$.  The receiver loads denoted by  $Z_L$ for the coils of $Rx$ are set to  $R$  without requiring any calibration with a trade-off in the received power performance.   Transmitter and receiver diversity are provided by $N_R < T$ loaded coils in $Rx$ (the remaining ones as passive coils) and all transmitter coils with the same resonance voltage.}    
  
{\color{black} Voltage sources and coil currents in the network denoted by the column vectors  $\mathbf{V}_{\omega}$ and $\mathbf{I_{\omega}}$, respectively, satisfy $\mathbf{V}_{\omega} = \mathbf{M}_{\omega} \, \mathbf{I_{\omega}}  $ where $\mathbf{M_{\omega}}$ is mutual inductance matrix, $\mathbf{V}_{\omega}^{T} =\big[0  \, \, \hdots \, \, 0 \, \, V_{2,\, 1}\, \, V_{2,\, 2}\, \, \hdots  \, \, V_{i,j}   \hdots  \, \, V_{K, \, T}\big]$, $\omega$ is the operating frequency, $V_{i,j}$ denotes the voltage level in $j$th coil of $i$th user, $i \, \in [2, \, K]$,  $j \in [1, \, T]$  and $\lbrace . \rbrace^{T}$ denotes transpose.}   $\mathbf{M}_{\omega}$ has the elements  $\mathbf{M}_{\omega}(i,j) = \jmath \, \omega \, M(i,j)$ for $i \neq j$, $Z_i \, + \, Z_L$ for $i = j \in [1, \, N_R]$ and $Z_i$ for  $i = j \in [N_R \,+ \,1, N_{Tot}]$ where $Z_i = R \, + \, \jmath \, \omega L \, +  \, 1 \, / \, (\jmath \, \omega \, C_{s})$, $\jmath = \sqrt{-1}$ is the complex unity and $M(i,j) = M(j,i)$ is the mutual inductance between $i$th and $j$th coil. {\color{black}Next,  the network topology modulation is introduced.}

\section{Network Topology Modulation}
\label{stm}

{\color{black}The spatial grid topologies of each transmitter are changed to form varying  mutual couplings. A signal modulating voltage waveform is not required but only varying spatial patterns of actively resonating coils.} Therefore, the network is continuously transmitting power to the receiver by also embedding information into the {\color{black}spatial structure that the energy is transmitted, however, without any signal modulation complexity or energy reservation for data.} The network topology modulation is mainly realized by changing $\mathbf{M}_{\omega}^{-1}$  denoted by  {\color{black}$ \mathbf{\Gamma}_{\omega}$} which is easily calculated by using matrix identities as follows: 
\begin{align}
\begin{split}
 \mathbf{\Gamma}_{\omega}  = &  {\color{black} \mathbf{Q} \, \bigg(\alpha_{\omega} \, \mathbf{\Lambda} \, +  \, \mathbf{I}  \,  +  \, \beta_{\omega} \,\sum_{i = 1}^{N_R} \vec{\mathbf{q}}_{i} \, \vec{\mathbf{q}}_{i}^H  \, \mathbf{\Lambda} \bigg)^{-1} \mathbf{Q}^H \, \mathbf{G}_{\omega}^{-1}}   \label{eq3}  
\end{split} 
\end{align}
where the elements  $\mathbf{M}(i,j)$ of {\color{black} $ \mathbf{M}= \mathbf{Q} \, \mathbf{\Lambda} \, \mathbf{Q}^H$} are equal to $M(i,j)$ for $i \neq j$ and $L$ otherwise, diagonal $\mathbf{\Lambda}$ includes eigenvalues, $\mathbf{Q} \, \mathbf{Q}^H = \mathbf{Q}^H \, \mathbf{Q} = \mathbf{I}$,  $\mathbf{G}_{\omega}^{-1} \equiv (\jmath \, \omega)^{-1}\big( \alpha_{\omega} \mathbf{I} + \beta_{\omega} \mathbf{I}_L \big)$,  {\color{black} $\mathbf{I} $ is the identity matrix}, $\vec{\mathbf{q}}_{i}^H$  is the $i$th row of $\mathbf{Q}$,  $\mathbf{I}_L(i,j) = 1$  for $i = j \in [1, N_{R}]$ and zero otherwise,  $\alpha_{\omega} = \jmath \, \omega^2 \,C_s \, / \, \varsigma_{\omega}$,  $\beta_{\omega} = - \jmath \,\omega^3 \,C_s^2 \,Z_L \, / \, \big( \varsigma_{\omega}^2 \, + \, \varsigma_{\omega}\, C_s \, \omega \, Z_L \big)$ and $\varsigma_{\omega} = (C_s \, \omega\, R \,-\, \jmath)$. The proof is given in Appendix \ref{proof1}.

It is observed in simulation studies that different network topologies provide different $\mathbf{\Lambda}$ allowing to realize \textit{eigenvalue modulation}. Eigenvalue modulation symbols $\mathbf{\Lambda}_k$ are either pre-determined for each $k \in [1, N_c]$  by using pilot training phase or they are estimated based on the preliminary knowledge of spatial modulation patterns and physical size of LAN topology. {\color{black} It is observed that $\mathbf{\Lambda}$ is robust to angular disorientations while the detailed analysis is left as a future work. In the following discussions, four different modulation methods are proposed in terms of oscillation frequencies of each coil and time sharing.}  

\setlength\tabcolsep{3 pt}    
\newcolumntype{M}[1]{>{\centering\arraybackslash}m{#1}}
\renewcommand{\arraystretch}{1.45}
\begin{table}[t!]
\caption{Network Topology Modulation Types and Properties}
\begin{center}
\scriptsize
\begin{tabular}{|c|c|c|c||c|c|c|c|}
\hline
ID    & Frequency & Time & $N_c$ & ID    & Frequency & Time & $N_c$\\ 
\hline
\hline
1   & $\omega_k$ for $k$th coil & Concurrent & $n^{K-1}$ & 3   &  $S_{\omega}$ (for all)  & Concurrent & $n^{K-1}$ \\    
\hline
2   & $\omega_k$ for $k$th coil & TDMA & $n$ & 4  & $S_{\omega}$ (for all)   & TDMA & $n$\\    
\hline
\end{tabular}
\end{center}
\label{tab1}  
\end{table}
\renewcommand{\arraystretch}{1}
\setlength\tabcolsep{6 pt}    

\subsection{Modulation with Orthogonal Time-Frequency Sharing}
 
Assume that each coil with the index $t$ in a grid transmits power at a different frequency $\omega_t$ for $t \in [1, T]$. Furthermore, {\color{black}voltage level $V$ is set to unity with equal phase for all coils for simplicity.} Then, the current at $\omega_t$ in $i$th coil of the receiver, i.e., $\mathbf{I_{\omega_t}}(i)$, is given as follows:
\begin{equation}
\label{eq5a}
 \mathbf{I_{\omega_t}}(i) = \sum_{j = 1}^{K-1} \mathbf{\Gamma_{\omega_t}}(i, t \, + j \, \, T) \, +  \, n_{t, i}
 \end{equation}  
where $n_{t, i}$ is assumed to be independent and identically distributed complex valued additive white Gaussian noise (AWGN) with variance of $\sigma_{t, i}^2$ and $E \lbrace n_{t, k} \, n_{m, l}^{*}\rbrace = 0$ for $m \neq t$ or $k \neq l$.  Moreover, if time-division multiple access (TDMA) channel sharing is utilized, then the induced current at a time slot is due to some specific $Tx_j$  for $j \in [1, \, K\, - \,1]$  and $\mathbf{I_{\omega_t}}(i)$ becomes equal to the following:
\begin{equation}
\label{eq5}
 \mathbf{I_{\omega_t}}(i) = \mathbf{\Gamma_{\omega_t}^j}(i, t \, + j \, \, T) \, +  \, n_{t, i}
\end{equation}
where $\mathbf{\Gamma^j_{\omega_t}}$ is obtained by deleting the rows and columns of 
$\mathbf{\Gamma_{\omega_t}}$  with the indices {\color{black} not in} $[j\,T \, + \, 1, (j \,+ \,1)\, T]$.
\subsection{Modulation with Network Topology Diversity}

If all the coils utilize the same frequency set in parallel, i.e., $ S_{\omega} =\lbrace \omega_s:  s \in [1, N_{\omega}] \rbrace $, where $N_{\omega}$ is the number of concurrent frequencies then,  the transmitter diversity  is realized and the received current is given as follows:
\begin{equation}
\label{eq6}
\mathbf{I_{\omega_s}}(i) = \sum_{t = T +1}^{N_{Tot}} \mathbf{\Gamma_{\omega_s}}(i, t) \, +  \, n_{s, i}
 \end{equation}  
{\color{black}If TDMA is utilized then, $\mathbf{I_{\omega_s}}(i)$ due to $Tx_j$ is as follows:}
\begin{equation}
\label{eq7}
 \mathbf{I_{\omega_s}}(i) = \sum_{t = 1}^{T} \mathbf{\Gamma_{\omega_s}^{j}}(i, t + j \, T)\, +  \, n_{s, i}
\end{equation} 
{\color{black} Assume that a set of symbols consisting of $n$ different topologies for each transceiver is utilized.  The total number of different symbol combinations excluding $Rx$ is denoted by $N_c$ and $i$th network symbol is denoted by $\mathbf{s}_i$.} {\color{black}The possible set of modulation types indexed with $Mod_{k}$ for $k \in [1, \, 4]$ defined in (\ref{eq5a}),  (\ref{eq5}), (\ref{eq6}) and (\ref{eq7}), respectively, and their properties are shown in Table \ref{tab1}.} $Mod_{2}$ and $Mod_4$ have the constellation size of $N_c = n$ while it becomes  $n^{K -1}$ for $Mod_1$ and $Mod_3$. The best time-frequency diversity is achieved with $Mod_3$ given in (\ref{eq6}) which is discussed in the following sections. 
{\color{black} The computational complexity of the proposed modulation scheme includes the calculation of total transmit power at the sources and normalizing the power resulting in $N_S \, N_\omega$ multiplications and $T^2$ additions where $N_S$ denotes the number of  random frequency sets ($S_\omega$) to include more diverse frequency effects. Next, demodulation mechanisms are discussed.} 

\section{Network Topology Demodulation}
 \label{stdm}

The demodulation is achieved by finding the spatial modulation matrix $\mathbf{\Gamma_{\omega}}$ for different symbols and at different frequencies, and then comparing with the received current at varying frequencies. Furthermore, the symbols are defined for the whole network {\color{black} by solving  MAC sharing problem} without requiring synchronization or time-frequency resource sharing. {\color{black} If we assume that $Mod_3$ is used, the network topology symbol} has the index $k$ for $k \in [1, N_c]$ and $\omega$ is the power transmission frequency then,  the measured noisy current denoted by $\mathbf{\tilde{I}_{\omega}}(i)$ {\color{black}at the} $i$th {\color{black}coil} receiver is found by combining (\ref{eq3}) and (\ref{eq6}) as follows:
\begin{equation}
\label{demod1}
\mathbf{\tilde{I}_{\omega}}(i) = \mathbf{I}_{\omega}(i) +  \, n_{s, i} = \vec{\mathbf{q}}_{i, k}^H \, \mathbf{\Upsilon}_{\omega, k}   \, \vec{\mathbf{q}}_{\Sigma, k} \, +  \, n_{s, i}
\end{equation}
where $ \vec{\mathbf{q}}_{i, k}^H$ is the $i$th row of $\mathbf{Q}$ in (\ref{eq3}),  $\vec{\mathbf{q}}_{\Sigma, k} = \sum_{j = T+1}^{N_{Tot}} \vec{\mathbf{q}}_{j, k}$,  $\mathbf{\Upsilon}_{\omega, k} = \alpha_{\omega}  \, (\jmath \, \omega)^{-1} \big(\mathbf{\Theta}_{\omega, k}  \,  +  \, \beta_{\omega} \,\sum_{i = 1}^{N_R} \mathbf{\Psi}_{i,k } \big)^{-1}$, $\mathbf{\Psi}_{i,k} = \vec{\mathbf{q}}_{i, \, k} \, \vec{\mathbf{q}}_{i, \, k}^H  \, \mathbf{\Lambda}_k$ and $\mathbf{\Theta}_{\omega, k} = \alpha_{\omega} \, \mathbf{\Lambda}_k \, +  \, \mathbf{I}$.  The proof is given in Appendix \ref{proof2}. It is further simplified with $N_{R} = 1$, $N_{Tot} = 1 + (K\,-\,1)\,T$ and by using perturbation equality $(\mathbf{I}  \, + a \,\vec{x} \,\vec{y}^H )^{-1} = \mathbf{I} \, + \, b \,  \vec{x} \,\vec{y}^H$ where $b = -a \, / \, (1 \, + \, a \, \vec{y}^H \, \vec{x})$ and the proof is given in  Appendix \ref{proof3}.  Then, $\mathbf{\Upsilon}_{\omega, k}$ becomes as follows:
\begin{equation}
\label{demoduoneuser2}
\begin{aligned}
\mathbf{\Upsilon}_{\omega, k} \, = \,   \frac{\alpha_{\omega} \, \mathbf{\Theta}_{\omega, k}^{-1}}{ \jmath \, \omega}  
   - \, \frac{\alpha_{\omega} \, \beta_{\omega} \, \mathbf{\Theta}_{\omega, k}^{-1}   \mathbf{\Psi}_{1,k}  \, \mathbf{\Theta}_{\omega, k}^{-1} \, (\jmath \, \omega)^{-1} }{  1 \,+\, \beta_{\omega} \,\vec{\mathbf{q}}_{1, k}^H \, \mathbf{\Lambda}_k \,  \mathbf{\Theta}_{\omega, k}^{-1} \vec{\mathbf{q}}_{1, k}}  & 
\end{aligned}
\end{equation}
{\color{black} Then, inserting into (\ref{demod1}) and dropping $i = 1$, it becomes $\tilde{I}_{\omega_s}  = \vec{\mathbf{\upsilon}}_{\omega_s, k}^T \,    \big( \vec{\mathbf{q}}_{1, k}^{*}    \otimes \vec{\mathbf{q}}_{\Sigma, k} \big)$ where $\omega = \omega_s$, $\otimes$ is the Kronecker product and  $\vec{\mathbf{\upsilon}}_{\omega, k}(t)$ for $t \in [1, \, N_{Tot}]$ is given by the following:}   
\begin{equation}
\label{demoduoneuser3}
\begin{aligned}
\vec{\mathbf{\upsilon}}_{\omega, k}(t) \, = \,  \frac{\alpha_{\omega}}{\jmath \, \omega} \bigg(    \varphi_{\omega, k, t}  -      \frac{   \sum_{j = 1}^{N_{Tot}} \varrho_{\omega, k, j}   \,   \lambda_{k,t} \, \varphi_{\omega, k, t}    }{  1 \,+\,    \sum_{j = 1}^{N_{Tot}}  \lambda_{k,j} \, \varrho_{\omega, k, j}   }  \bigg)& 
\end{aligned}
\end{equation}  
where {\color{black}$\varrho_{\omega, k, j} \equiv \beta_{\omega}\, \vert q_{k,j}\vert^2   \, \varphi_{\omega, k, j}$, $\varphi_{\omega, k, t} \equiv (\alpha_{\omega}  \lambda_{k,t}  \, + \,1)^{-1}$,} $\lambda_{k,t} \equiv \mathbf{\Lambda}_k (t,t)$ and $q_{k,t} \equiv  \vec{\mathbf{q}}_{1, k}(t)$. If it is assumed that eigenvalues are close to each other then, $\vec{\mathbf{\upsilon}}_{\omega, k}(t)$ is approximated by dropping the $q_{k,j}^2$ terms due to the unitary magnitude property of  $ \vec{\mathbf{q}}_{1, k}$. {\color{black} Then, the received  current vector $\mathbf{\tilde{I}}_{S} = [ \tilde{I}_{\omega_1} \, ... \, \tilde{I}_{\omega_{N_{\omega}}} ]^T$} is  approximated by $\mathbf{\tilde{I}}_{S} \approx \eta_{k} \mathbf{\Upsilon}_{k} \, \vec{\mathbf{q}}_{0,k} \, + \, \mathbf{n}_{S}$ where {\color{black} $\mathbf{\Upsilon}_{k}^T \equiv \left[ \vec{\mathbf{\upsilon}}_{\omega_1, k} \, \hdots \, \vec{\mathbf{\upsilon}}_{w_{N_{\omega}}, k} \right]$,} $\mathbf{n}_{S} = [n_{\omega_1} \, ... \, n_{\omega_{N_{\omega}}} ]^T$ is the complex receiver noise vector  having $N_0/2$ spectral density for real and imaginary parts at each frequency, $\vec{\mathbf{q}}_{0,k} \equiv \vec{\mathbf{q}}_{1, k}^{*}    \otimes \vec{\mathbf{q}}_{\Sigma, k}$  {\color{black}satisfying $\sum_{t=1}^{N_{Tot}} \vec{\mathbf{q}}_{0,k}(t) = 0$ and  $\eta_k$  equalizes the received or transmitted power for the symbol.}  

\begin{figure}[!t]
\vspace{0.05in}
\centering
\includegraphics[width=2.0in]{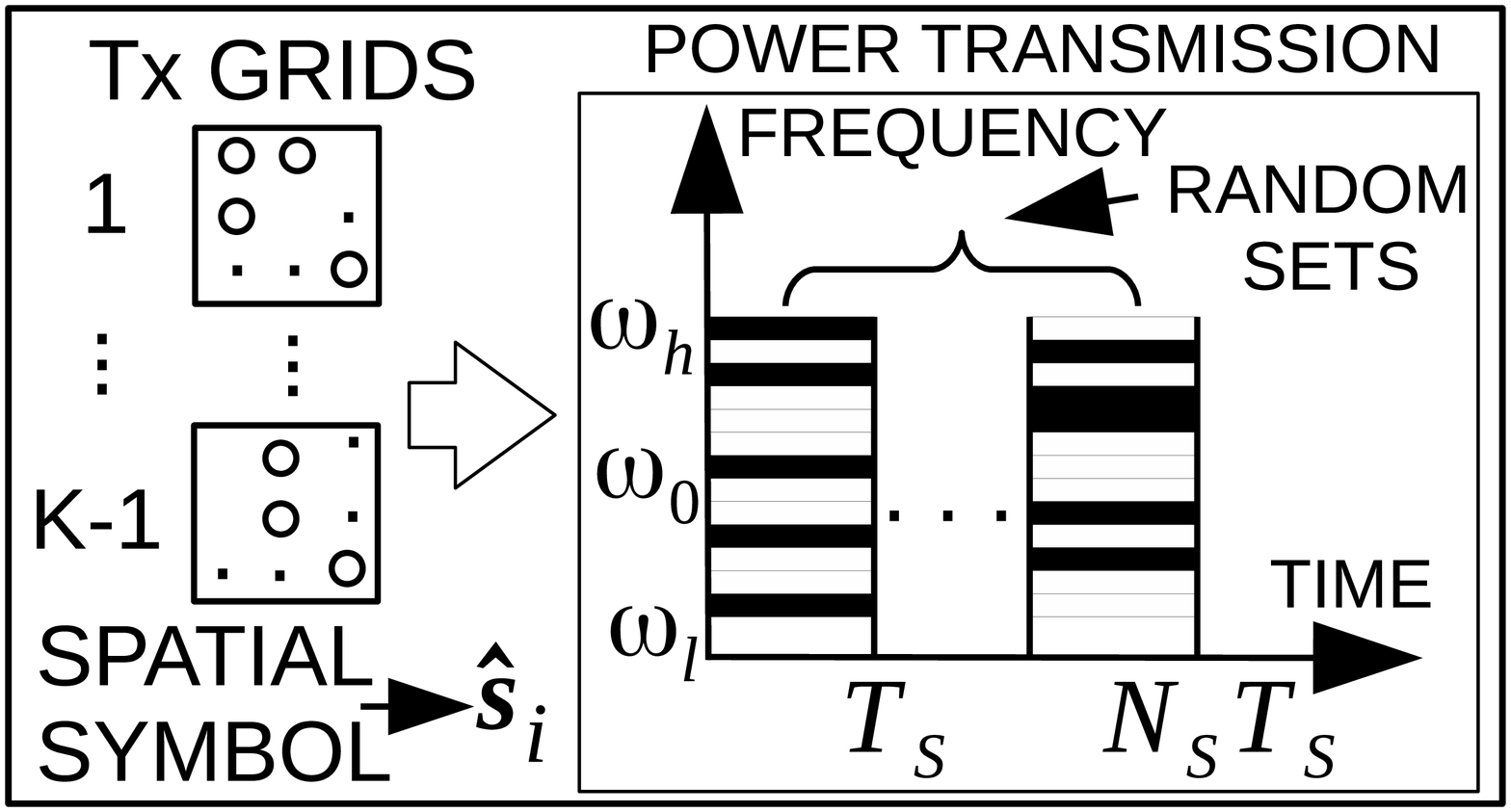} \\
(a) \\
\vspace{0.03in}
\includegraphics[width=2.8in]{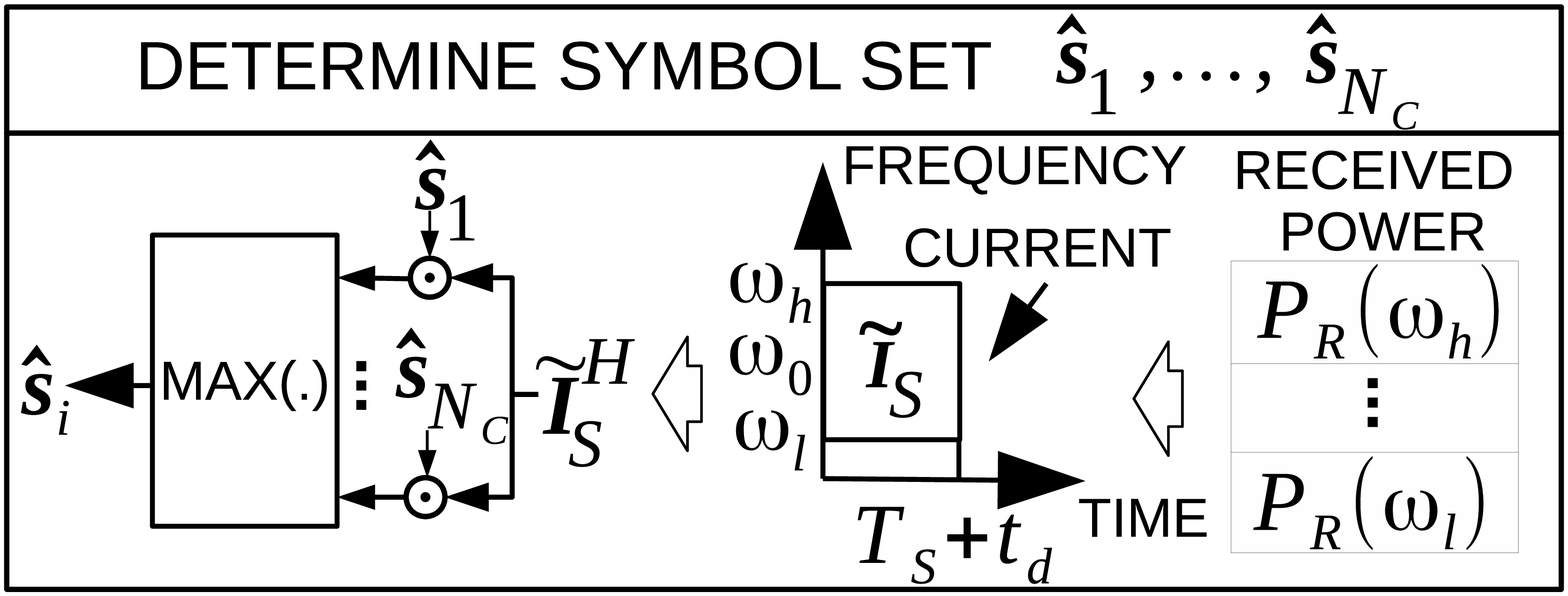} \\
(b)
\vspace{-0.08in}
\caption{ (a) {\color{black}Network topology modulation,} and (b) demodulation mechanism for pilot aided case with full knowledge of the transmitted symbol set.}
\label{fig5}
\end{figure}

There are two different {\color{black}demodulation methods} with respect to the amount of knowledge about   $\eta_{k} \, \mathbf{\Upsilon}_{k} \, \vec{\mathbf{q}}_{0,k}$ for each symbol $\mathbf{s}_k$ such that it is either perfectly known by pilot aided transmission or $\mathbf{\Upsilon}_{k}$  is estimated based on pre-calculation for the communication ranges without any information about $\eta_{k} \, \vec{\mathbf{q}}_{0,k}$.  It is observed in Section \ref{numersim} that the eigenvalues in $\mathbf{\Lambda}_k$  have  low variances for varying positions and orientations, and they are specific {\color{black}for each symbol $\mathbf{s}_k$. Demodulation schemes are realized by utilizing this uniqueness as shown in the following.}

\subsection{Pilot Aided Perfect Knowledge of Symbol Set} 
The receiver is assumed to have the full knowledge of $\eta_k$, $\mathbf{\Upsilon}_{k}$ and $\vec{\mathbf{q}}_{0,k}$ for any $k$ with the aid of pilot symbols. {\color{black}The modulation schemes  are shown in Fig. \ref{fig5}(a) where $N_{S} = 1$ in pilot training case. The demodulation mechanism is shown in Fig. \ref{fig5}(b). Each symbol $\mathbf{s}_k$ is the vector $\eta_k \,\mathbf{\Upsilon}_{k} \, \vec{\mathbf{q}}_{0,k}$ with  period $T_s$, transmission delay $t_d$ and $\mathbf{n}_{S}$ is the AWGN component with independent and identically distributed elements having variance $\sigma_{n_{\omega_s}} = N_0$ for $s \in [1, N_{\omega}]$.  The optimum estimation is achieved with maximum likelihood (ML) decision using the distance metric, i.e.,  $0.5 \, \mathbf{s}_k^H \,\mathbf{s}_k -  \mathbf{\tilde{I}}_{S}^H \, \mathbf{s}_k $, and choosing the minimum based on decision regions $D_k$ for each $\mathbf{s}_k$ \cite{proakis}. The computational complexity includes finding the maximum of the correlations and the term $\mathcal{O} (N_\omega \, n^{K-1})$ for $Mod_3$  due to $N_\omega$ multiplications and additions in $\mathbf{\tilde{I}}_{S}^H \, \mathbf{s}_k$ with $N_S = 1$. } {\color{black} Transmit powers are adapted with $\eta_k$ to equalize total received power  for each  $k$ for simplicity.}   {\color{black}Then, with equally probable symbols, the probability of error per transfer is given by $P_e = 1-\frac{1}{N_c}\sum_{m  =1}^{N_c} \int_{D_{m}} f_{\mathbf{\tilde{I}}_{S}}(\mathbf{\tilde{I}}_{S} \vert \mathbf{s}_m)\, \mbox{d}\mathbf{\tilde{I}}_{S}$ where     $f_{\mathbf{\tilde{I}}_{S}}(\mathbf{\tilde{I}}_{S} \vert \mathbf{s}_m)$ is approximated by $(\pi\, N_0)^{-\frac{N_{\omega}}{2}    } \mbox{exp}\big(- \frac{1}{N_0} \sum_{j=1}^{N_{\omega}}\vert\tilde{I}_{\omega_j} - s_{mj}\vert^2 \big)$ and $s_{mj}$ is the $j$th element of $\mathbf{s}_m$. The upper bound is given by $P_e \leq (0.5 \, / \, N_c) \, \sum_{i=1}^{N_c} \sum_{k=1, k \neq i}^{N_c} \mbox{erfc}(d_{ik} \, / \, 2 \sqrt{2 \, N_0})$ where $d_{ik}$ is the distance between $\mathbf{s}_i$  and $\mathbf{s}_k$, and $\mbox{erfc}(.)$ is the complementary error function.}
  
\subsection{Pre-computation based Demodulation} 

\begin{figure}[!t]
\vspace{0.05in}
\centering
\includegraphics[width=2.5in]{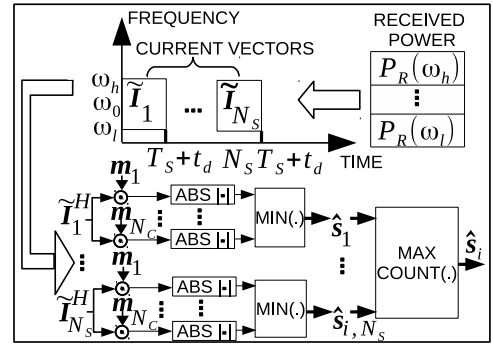} \\
\vspace{-0.05in}
\caption{{\color{black}Network topology demodulation without pilot training.}}
\label{fig6}
\end{figure}

It is assumed that there is no knowledge about the symbol set or  $\eta_{k} \, \vec{\mathbf{q}}_{0,k}$. An average  $\overline{\mathbf{\Upsilon}}_{k}$ is pre-calculated based on the observation of the small variance of $\mathbf{\Lambda}_k$ and by considering the approximation in (\ref{demoduoneuser3}) without $q_{k,j}^2$ terms for the estimated range and orientations of the transmitters. The proposed demodulation scheme is shown in Fig. \ref{fig6}. {\color{black}The transmit power for  $\mathbf{s}_m$ is equalized with $\eta_{m}$ instead of the receiver power since there is no feedback. Then, the estimation of $\mathbf{s}_m$ is achieved by finding the index $k$ minimizing {\color{black}$\vert \sum_{t=1}^{N_{Tot}} \widehat{\eta}_{k} \, \widehat{\mathbf{q}}_{0,k,m}(t) \vert^2$} where $\widehat{\eta}_{k} \, \widehat{\mathbf{q}}_{0,k,m} = \overline{\mathbf{\Upsilon}}_{k}^{\dagger} \, \mathbf{\tilde{I}}_{S}$, $\widehat{\eta}_{k} \, \widehat{\mathbf{q}}_{0,k,m}$ denotes parameter estimation regarding symbol $\mathbf{s}_k$ when symbol $\mathbf{s}_m$ is transmitted, $\mathbf{\tilde{I}}_{S} = \eta_{m} \mathbf{\Upsilon}_{m} \, \vec{\mathbf{q}}_{0,m} \, + \, \mathbf{n}_{S}$ and $\overline{\mathbf{\Upsilon}}_{k}^{\dagger}$ denotes Moore-Penrose pseudoinverse of $\overline{\mathbf{\Upsilon}}_{k}$.   Then, $\vert \sum_{t=1}^{N_{Tot}} \widehat{\eta}_{k} \, \widehat{\mathbf{q}}_{0,k,m}(t) \vert^2 $ equals to $  \vert   z_{k, m} \vert^2$ where  $z_{k, m} \equiv g_{k,m} \, + \,\mathbf{m}_{k}^T \, \mathbf{n}_{S} $, $g_{k,m} \equiv  \mathbf{1}^T \,  \overline{\mathbf{\Upsilon}}_{k}^{\dagger} \, \mathbf{\Upsilon}_{m} \,  \eta_{m} \, \vec{\mathbf{q}}_{0,m}$,  $ \mathbf{m}_{k}^T \equiv  \mathbf{1}^T \,  \overline{\mathbf{\Upsilon}}_{k}^{\dagger}$ and $\mathbf{1}^T$ is the row vector of all ones. $ \chi^2_{k,m}(1) \equiv \vert   z_{k, m} \vert^2$ is a generalized non-central chi-square distribution with the number of degrees of freedom of one where $z_{k, m}$ is complex Gaussian random variable with the expectation  $E \lbrace z_{k, m} \rbrace = g_{k,m}$ and variance $\mbox{Var} \lbrace z_{k, m} \rbrace = N_0 \Vert \mathbf{m}_{k}   \Vert_{2}^2$, and $\Vert .\Vert_{2}$ denotes Euclidean norm.}  
If the probability $P(\chi^2_{m,m}(1) > \chi^2_{j,m}(1))$ is denoted by $P_e(m, j)$ then, the probability of error $P_e$ is bounded by $  (1 \, / \, N_c)  \, \sum_{i=1}^{N_c} \sum_{k=1, k \neq i}^{N_c} P_e(i, k)$.  The probability distribution of the difference of two correlated non-central chi-square random variables is complex to formulate explicitly.  In addition,  {\color{black}random sets of energy} transmission frequencies are utilized to improve detection performance by averaging the true decisions given for each set $ S_{\omega, i} =\lbrace \omega_{s,i}:  s \in [1, N_{\omega}] \rbrace$ for $i \in [1, N_S]$ as shown in Fig. \ref{fig6}.   Furthermore, the  noise is smaller in each detection procedure where $N_{\omega}$ different frequencies are utilized instead of $N_{\omega} \times N_{S}$.  {\color{black} The computational complexity includes finding  the minimums of $N_S$ different vectors of length $N_c$, finding the maximum of a vector of size $N_c$  and the term $\mathcal{O}(N_S \, N_\omega \, N_c) = \mathcal{O}  (N_S \, N_\omega \, n^{K-1})$ for $Mod_3$ due to $N_S  \, N_c \, ( N_\omega\, + \, 1)$ multiplications and  $N_S \, N_c\, N_\omega $ additions.}
 
\section{Numerical Simulations}
\label{numersim}

\begin{figure}[!t]
\vspace{0.05in}
\centering
\includegraphics[width=1.2in]{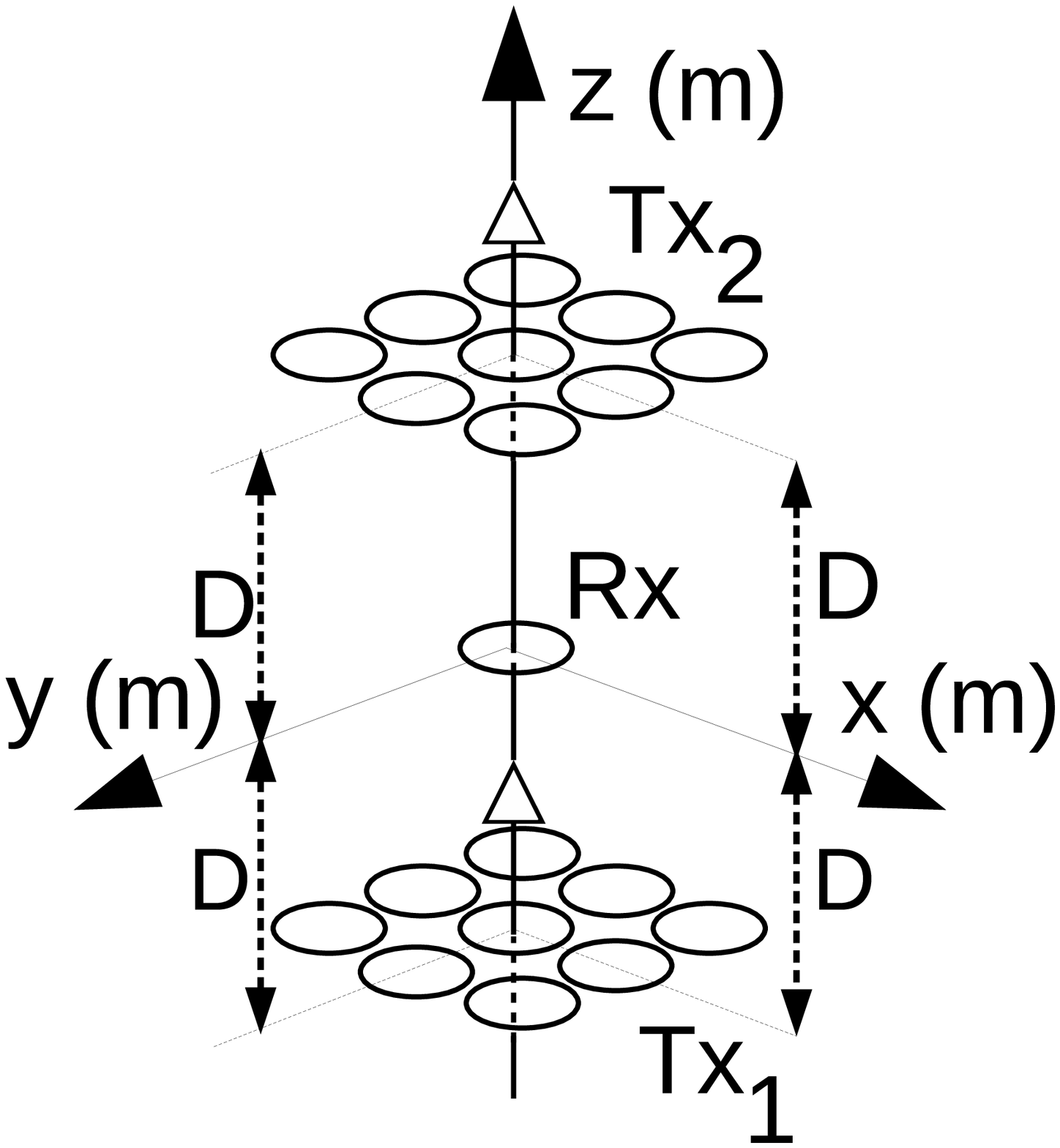} \includegraphics[width=1.45in]{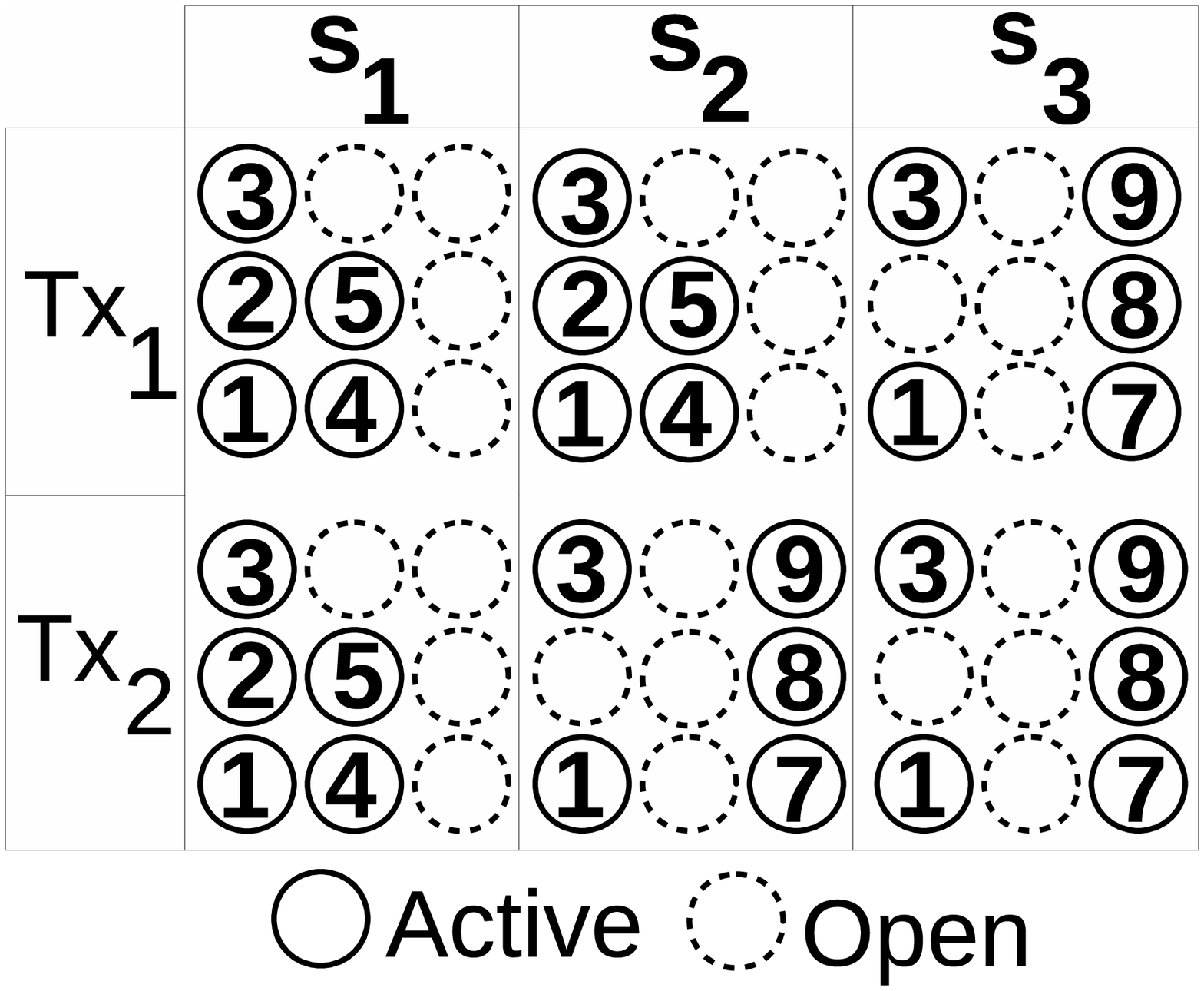} \\
(a) \hspace{1.1in}  (b) 
\vspace{-0.05in}
\caption{ (a) Two-user MAC topology with mirror symmetric positions of the parallel transmitters and (b) three different network topology symbols.}
\label{fig34}
\end{figure}

The proposed {\color{black}system} is simulated for {\color{black}two-user MAC scheme as a proof of concept.} It is {\color{black} quite} easy to extend to crowded IoT networks. The receiver is chosen to have {\color{black}a single coil to simplify the analysis.} The  orientations of the transmitter coils are chosen {\color{black}in parallel with $Rx$ with equal distances to simplify the analysis}  as shown in Fig. \ref{fig34}(a).   The positions of the coils  are represented in Cartesian coordinate system as $\mathbf{p}_1 = \big[ 0  \, \, \,  0  \,  \, \,  -D \big]^T$ and $\mathbf{p}_2 = \big[0 \, \, \, 0  \, \, \, D \big]^T$. Their equal normal vector is $\mathbf{\vec{n}} = [ 0 \, \, 0 \, \, 1]$.  {\color{black} $N$ and $T$ are set to three and five, respectively}. {\color{black}Modulation topology for a single coil is assumed to be one of two schemes}, i.e., either the coils with the indices $\lbrace 1,2,3,4,5 \rbrace$ or $\lbrace 1,3,7,8,9 \rbrace$ are active at any time. Therefore, there is a total of three different {\color{black}network topology symbols} as shown in Fig. \ref{fig34}(b). {\color{black}Single channel use transmits $\log_2 3$ bits of total MAC data.} $N_c$ can be easily increased, however, three symbols better clarify the proposed system.

\setlength\tabcolsep{3 pt}    
\newcolumntype{M}[1]{>{\centering\arraybackslash}m{#1}}
\renewcommand{\arraystretch}{1.35}
\begin{table}[t!]
\caption{Simulation Parameters}
\begin{center}
\scriptsize
\begin{tabular}{|m{1.75cm}|p{3.35cm}|p{3.1cm}| }
\hline
PARAM. & MEANING &    VALUE \\ 
\hline
\hline
$r, \, \Delta_c$ & The coil radius and inter-coil distance in the grids &    $1$ cm, $1$ mm \\
\hline
$R, \, L, \, C_s$ & The resistance, inductance and capacitance of a single coil unit &  $65 \, m\Omega, \, 84 \, $ nH, $3$ - $300$ nF  \\ 
\hline
$f_0$ & Operating resonance frequency &   $1$ MHz, $10$ MHz  \\
\hline
$\omega_l, \, \omega_h \, ( \times \, \omega_0^{-1})$ & Transmission frequency interval &   $\lbrace 0.92$ - $ 1.08 \rbrace$, $\lbrace 0.99$ - $1.01 \rbrace$ \\
\hline
$N_S$, $N_{\omega}$, $N_{c}$  & The number of random freq. sets, {\color{black}freqs.} in a set, and {\color{black}the} symbols & $\lbrace 1, \, 24\rbrace$, $\lbrace 24, \, 32\rbrace$, $3$  \\
\hline
$N_0$ & Noise power spectral density &   $10^{-21}, 10^{-16}, 10^{-12} $ W/Hz  \\
\hline
$\overline{P}_T$ & Average trans. power per symbol &   $1 \, / \, N_c = 1\,/ \,3$ mW  \\
\hline
\end{tabular}
\end{center}
\label{tab3}  
\end{table}
\renewcommand{\arraystretch}{1}
\setlength\tabcolsep{6 pt}  
 
{\color{black} The simulation parameters are shown in Table \ref{tab3}. The coil radius and the boundary distance between the coils are set to $r = 1$ cm and $\Delta_c = 1$ mm, respectively, so that a transceiver with a planar grid of coils has the area of $6.2 \times 6.2$ $\mbox{cm}^2$ to be easily attached to daily objects as shown in Fig. \ref{fig12}(a). The number of turns is  one for simplicity. $R$ is calculated by using  $ \pi \, r \, \rho  \, / \,( 2  \, w_c \, \delta)$ for square cross-section copper wire of width and height $w_c = 2$ mm  compatible with \cite{burhan3} and \cite{ref1} where the resistivity is $\rho= 1.72\,10^{-8} \,\Omega \, \mbox{m}$, skin depth is $\delta = \sqrt{\rho \,  / \, ( \pi \, \mu \, f_0)}$ and  $\mu = 4 \, \pi \, 10^{-7}$ $\mbox{H} \, / \, \mbox{m}$.  $L$ is equal to $\big( \mu \, l_c \, / \, (2\,\pi) \big) \big( \ln\big( l_c \, / \, w_c\big) \, + \,  0.5  \, + \, 0.447 \, w_c \, / \, l_c \big) \approx 84$ nH  where $l_c \, \equiv \, 2 \, \pi \, r$ \cite{rosa}.  The capacitance is $C_s = 1 \, / \, (\omega_0^2 \, L)  \approx  300$ nF or $3$ nF for $f_0 = 1$ MHz or $10$ MHz, respectively, with $f \leq 10 $ MHz to reduce parasitic effects \cite{burhan3, ref1}. Noise spectral density is simulated for $N_0 = \big[ 1 \, \, 10^5 \, \, 10^9 \big] \times \mbox{N}_{Th} $ to analyze worse cases than the thermal noise where $\mbox{N}_{Th} = 4 \, \kappa_B \,T_r \, R  \approx \, 10^{-21}$ W/Hz, $\kappa_B  \, = \, 1.38 \times 10^{-23}$ J/K and $T_r\, = \,  300$ K \cite{burhan3, ref1}.   Total transmit  power for $N_c$ symbols is set to $1$ mW in low power regime with an average $\overline{P}_T = 1 \, / \, N_c$ (mW)  for massive utilization in everyday objects.  Mutual inductance  calculation is based on the general model in \cite{burhan3} and \cite{babic-validity} including position, orientation and size dependency. The demodulation mechanisms with and  without any training are simulated next.} 

\subsection{Pilot Training}

The case with pilot training is simulated for $f = 1$ MHz, $(\omega_l, \omega_h) = (0.99, 1.01) \times \omega_0$, {\color{black} $N_{\omega} = 32$ and $N_{S} = 1$  due to perfectly known  symbol family allowing to decrease $N_S$.  The received power for each symbol is equalized to simplify the system and the analysis  where unequal transmit powers are used for each symbol with the average $\overline{P}_T = 1/ 3$ mW and distributed equally among $N_{\omega}$ frequencies.} 

\begin{figure}[!t]
\vspace{0.05in}
\centering
\includegraphics[width=2.8in]{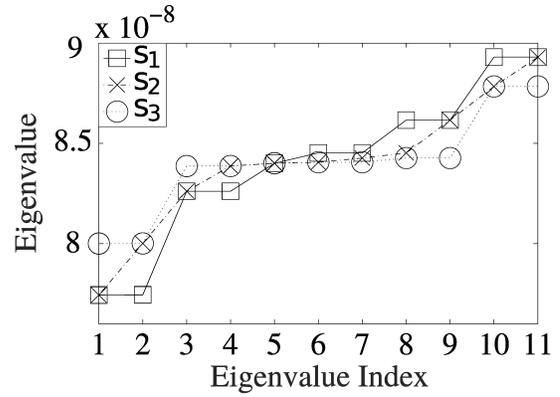} \\
(a) \\
\includegraphics[width=2.8in]{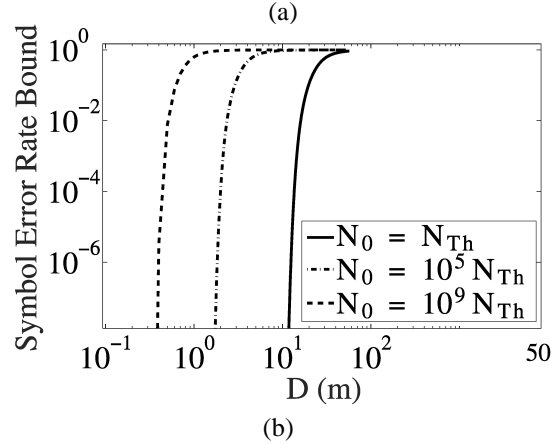} \\
(b)
\vspace{-0.05in}
\caption{ (a) {\color{black}Eigenvalues of $\mathbf{Q}$ for different symbols and (b) the upper bound on the symbol error probability for varying $D$ and noise levels.}}
\label{MatlabFig12}
\end{figure}

{\color{black}Eigenvalues are plotted for  varying distances reaching to $\approx$ 50 meters as shown in Fig. \ref{MatlabFig12}(a).  They are robust with respect to the distance  which is utilized in the next section to demodulate by directly using the eigenvalues. The upper bound on the probability of symbol error is shown in Fig. \ref{MatlabFig12}(b) where it drops to $10^{-8}$ at several meter distances including high-noise regimes allowing for reliable IoT communications. It is possible to realize a LAN having the range of tens of meters with low cost, planar and small coils with the potential to tune system performance by changing system parameters.}   
 
\subsection{Without Pilot Training}

{\color{black}The small variations of eigenvalues shown in Fig. \ref{MatlabFig12}(a) for varying distances are utilized to practically demodulate without any training or synchronization among the transmitters.}  The case is simulated for $f = 10$ MHz, $(\omega_l, \omega_h) = (0.92, 1.08) \times \omega_0$ and $N_{S} = N_{\omega} = 24$. $\overline{P}_T$ is  distributed among $N_{S} \times N_{\omega}$ frequencies equally while equalizing only the transmit power for each symbol without any receiver feedback.  

\begin{figure}[!t]
\vspace{0.06in}
\centering
\includegraphics[width=3.2in]{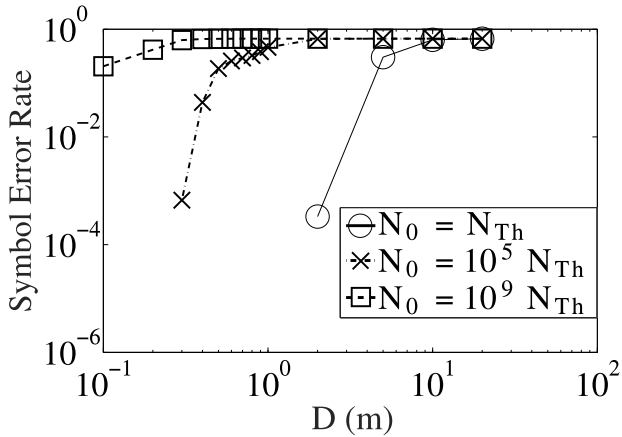}
\vspace{-0.06in}
\caption{{\color{black}Monte Carlo simulation of the probability of symbol error for demodulating without any pilot training for varying $D$ and noise levels.}}
\label{MatlabFig3}
\end{figure}

The calculation of error probability is realized through Monte Carlo simulations with $4 \times 10^{6}$ trials due to complexity of obtaining explicit expressions. Therefore, the only errors higher than $2.5 \times 10^{-6}$ are visible in the simulation results. It is observed in Fig. \ref{MatlabFig3} that the proposed system achieves reliable communication in several meters without any pilot training or receiver feedback. {\color{black} In comparison with far field RF based SWIPT systems, far field directive power beaming achieves several meters of ranges for indoor and outdoor applications achieving to transmit several mWs of power \cite{swiptrange}. However, high efficiency rectifiers require input powers between $0.5$ - $5$ mW  with much higher transmission powers although the radiative RF range is not limited. On the other hand, it is possible to achieve MI SWIPT with several mWs transmission power with the possibility to tune and further improve the range performance by optimizing coil dimensions, the number of coils, constellations and operating frequency,  and by utilizing passive waveguides or relays, e.g., enabling ranges reaching kilometers with larger radius coils \cite{ref1}.} The problems of synchronization, channel access and resource sharing  are solved by utilizing time-frequency resources together by all the transmitters and by only changing the network topology. 

\section{Open Issues and Future Work}
\label{openis}
{\color{black}
The following topics  promise to further improve the system:
\begin{itemize}
\item The computational complexity exponentially dependent on $K$ due to multi-user ML detection needs to be reduced with modern methods, e.g., sphere decoding, by analyzing the structures of $\eta_k$, $\mathbf{\Upsilon}_{k}$, $\vec{\mathbf{q}}_{0,k}$ and $\overline{\mathbf{\Upsilon}}_{k}^{\dagger}$.
\item Analysis and experimentation of performance for varying the following: types and numbers of constellations;  numbers, positions and orientations of coil grids; coil and grid dimensions; frequency sets, voltage sources and synchronization; passive relaying and waveguides.
\item The effects of intruder coils due to change in the network topology need to be analyzed in more detail and a protocol needs to be developed to detect the changes in the received data and the consumed powers in the coils.
\end{itemize}
}

\section{Conclusion}
\label{conclusion}

{\color{black}In this paper,  novel modulation and demodulation mechanisms are proposed with energy efficient, low cost and low  hardware complexity mechanisms for SWIPT in a wireless MI MAC network. The network topology of power transmitting coils is changed to modulate information symbols and practical demodulation schemes  are presented. The designed scheme is numerically simulated for two-user MAC topology with small size coils suitable to attach to daily objects. The spatial modulation method provides reliable  LAN performances with tens of meters of ranges and low power operation of mW level transmit powers.  SWIPT obtained with MIC networks is future promising for IoT applications and MACs with energy efficient, simple and low cost solutions.}

\appendices

\section{}
\label{proof1}
$ \mathbf{M}_{\omega}$  is equal to $\jmath \, \omega \,\mathbf{M} \, + \, R \, \mathbf{I} \,  + \, Z_L \mathbf{I}_L  \,  +  \mathbf{I} \, / \, (\jmath \, \omega \, C)$. It is converted to $(\jmath \, \omega) \, (\mathbf{M} \, + \, \mathbf{E}_{\omega} ) $   where $\mathbf{E}_{\omega} = R \, \mathbf{I} \, / \, (\jmath \, \omega) \,  + \, Z_L \mathbf{I}_L  \, / \, (\jmath \, \omega) \, -  \mathbf{I} \, / \, (\omega^2 \, C)$. Then, $ \mathbf{M}_{\omega}^{-1}$ is given by $(\mathbf{M} \, + \, \mathbf{E}_{\omega} )^{-1}  \, / \, (\jmath \, \omega)  = (\mathbf{I} \,  + \, \mathbf{E}_{\omega}^{-1}  \, \mathbf{M})^{-1}\,\mathbf{E}_{\omega}^{-1}\, / \, (\jmath \, \omega)  $. Furthermore, $\mathbf{E}_{\omega}^{-1}$ is simplified as $  \alpha_{\omega} \mathbf{I} + \beta_{\omega} \mathbf{I}_L$. The inverse $ \mathbf{\Gamma}_{\omega} = \mathbf{M}_{\omega}^{-1} $ is found by inserting the simplification of $\mathbf{E}_{\omega}^{-1}$   as follows: 
\begin{eqnarray}
 \label{app_eq1}
\mathbf{\Gamma}_{\omega}  & \overset{1}{=} &  ( \alpha_{\omega} \, \mathbf{M} \,  + \mathbf{I} \,  + \,  \, \beta_{\omega} \mathbf{I}_L \, \mathbf{M})^{-1}\,\mathbf{G}_{\omega}^{-1} \,\,\,\,\,\, \nonumber \\ 
& \overset{2}{=} &  ( \alpha_{\omega} \,  \mathbf{Q} \, \mathbf{\Lambda} \, \mathbf{Q}^H \,  + \mathbf{I} \,  + \,  \, \beta_{\omega} \mathbf{I}_L \,  \mathbf{Q} \, \mathbf{\Lambda} \, \mathbf{Q}^H)^{-1}\,\mathbf{G}_{\omega}^{-1} \,\,\,\,\,\, \nonumber \\ 
&  \overset{3}{=} &   \mathbf{Q} \, \bigg(\alpha_{\omega} \, \mathbf{\Lambda} \, +  \, \mathbf{I}  \,  +  \, \beta_{\omega} \, \mathbf{Q}^{-1}  \mathbf{I}_L \, \mathbf{Q} \,   \mathbf{\Lambda} \bigg)^{-1} \mathbf{Q}^H \, \mathbf{G}_{\omega}^{-1} \,\,\,\,\,\, \nonumber \\
& \overset{4}{=} &   \mathbf{Q} \, \bigg(\alpha_{\omega} \, \mathbf{\Lambda} \, +  \, \mathbf{I}  \,  +  \, \beta_{\omega} \,\sum_{i = 1}^{N_R} \vec{\mathbf{q}}_{i} \, \vec{\mathbf{q}}_{i}^H  \, \mathbf{\Lambda} \bigg)^{-1} \mathbf{Q}^H \, \mathbf{G}_{\omega}^{-1} \,\,\,\,\,\, \,\,\,\,\,\,
\end{eqnarray}
where  $\mathbf{G}_{\omega}^{-1} \equiv (\jmath \, \omega)^{-1} \mathbf{E}_{\omega}^{-1}$ and $\vec{\mathbf{q}}_{i}^H$  is the $i$th row of $\mathbf{Q}$. The simplification while passing from $\overset{3}{=}$ to $\overset{4}{=}$ is realized by using the equality $\mathbf{I}_L \, \mathbf{Q} = \mathbf{Q} \, \sum_{i = 1}^{N_R} \vec{\mathbf{q}}_{i} \, \vec{\mathbf{q}}_{i}^H$ which can be obtained by using simple matrix calculations.

\section{}
\label{proof2}
The expression $\mathbf{\Gamma}_{\omega} \mathbf{V_{\omega}}$ to obtain the received current is simplified by using  (\ref{app_eq1})  as follows: 
\begin{eqnarray}
 \mathbf{\Gamma}_{\omega} \mathbf{V_{\omega}}  & \overset{1}{=}  &   \mathbf{Q} \, \big(\mathbf{\Theta}_{\omega, k}  \,  +  \, \beta_{\omega} \,\sum_{i = 1}^{N_R} \mathbf{\Psi}_{i,k } \big)^{-1} \mathbf{Q}^H \, \mathbf{G}_{\omega}^{-1} \mathbf{V_{\omega}} \nonumber  \\
 & \overset{2}{=}  &  \frac{1}{\jmath \, \omega} \mathbf{Q} \, \big(\mathbf{\Theta}_{\omega, k}  \,  +  \, \beta_{\omega} \,\sum_{i = 1}^{N_R} \mathbf{\Psi}_{i,k } \big)^{-1} \mathbf{Q}^H \, \nonumber  \\ 
&& ( \alpha_{\omega} \mathbf{I} + \beta_{\omega} \mathbf{I}_L) \mathbf{V_{\omega}} \nonumber \\
 & \overset{3}{=}  &  \frac{ \alpha_{\omega}}{\jmath \, \omega} \mathbf{Q} \, \big(\mathbf{\Theta}_{\omega, k}  \,  +  \, \beta_{\omega} \,\sum_{i = 1}^{N_R} \mathbf{\Psi}_{i,k } \big)^{-1} \mathbf{Q}^H \, \mathbf{V_{\omega}} \nonumber   \\
& \overset{4}{=}  &  \frac{ \alpha_{\omega}}{\jmath \, \omega} \mathbf{Q} \, \big(\mathbf{\Theta}_{\omega, k}  \,  +  \, \beta_{\omega} \,\sum_{i = 1}^{N_R} \mathbf{\Psi}_{i,k } \big)^{-1} \nonumber \\ 
&& \sum_{j = T+1}^{N_{Tot}} \vec{\mathbf{q}}_{j, k} \, V(j,\omega)
\label{app_eq2}
\end{eqnarray} 
where  it is assumed that $k^{th}$ symbol is transmitted, $\mathbf{V_{\omega}}(t) = V(t, \omega)$ is the oscillating voltage for the $t^{th}$ coil at $\omega$ for $t \in [T+1, N_{Tot}]$ and zero otherwise, the equality  $\overset{3}{=}$ is obtained by using $\mathbf{I}_L \mathbf{V_{\omega}} = 0$, and $\mathbf{Q}^H \, \mathbf{V_{\omega}} = \sum_{j = T+1}^{N_{Tot}} \vec{\mathbf{q}}_{j, k} \, V(j,\omega) \equiv \vec{\mathbf{q}}_{\Sigma, k}$ is utilized in the equality $\overset{4}{=}$. Then, the current at $i^{th}$ receiver coil is found by replacing $\mathbf{Q}$ with  $\vec{\mathbf{q}}_{i, k}^H$.

\section{} 
\label{proof3}
$\mathbf{I}_{\omega}(i)$ is calculated by using (\ref{app_eq2}) as follows:   
\begin{eqnarray}
 \mathbf{I}_{\omega}(i)   & \overset{1}{=}  &   \frac{ \alpha_{\omega}}{\jmath \, \omega} \, \vec{\mathbf{q}}_{i, k}^H \, \big(\mathbf{\Theta}_{\omega, k}  \,  +  \, \beta_{\omega} \,\sum_{i = 1}^{N_R} \mathbf{\Psi}_{i,k }  \big)^{-1} \, \vec{\mathbf{q}}_{\Sigma, k}  \nonumber \\
 & \overset{2}{=}  &   \frac{ \alpha_{\omega}}{\jmath \, \omega}\,  \vec{\mathbf{q}}_{i, k}^H \,  \big(\mathbf{I} \,  +  \, \beta_{\omega} \,\mathbf{\Theta}_{\omega, k}^{-1} \, \mathbf{A}_k\big)^{-1}  \mathbf{\Theta}_{\omega, k}^{-1} \,\vec{\mathbf{q}}_{\Sigma, k}   \,\,\,\,\,\,\,\,\,\,
\label{app_eq3}
\end{eqnarray} 
where  $\mathbf{A}_k = \sum_{i = 1}^{N_R} \mathbf{\Psi}_{i,k }  = \sum_{i = 1}^{N_R} \vec{\mathbf{q}}_{i, \, k} \, \vec{\mathbf{q}}_{i, \, k}^H  \, \mathbf{\Lambda}_k$. If $N_R = 1$, then the perturbation equality, i.e., $(\mathbf{I}  \, + a \,\vec{x} \,\vec{y}^H )^{-1} = \mathbf{I} \, + \, b \,  \vec{x} \,\vec{y}^H$ with $b = -a \, / \, (1 \, + \, a \, \vec{y}^H \, \vec{x})$, is utilized where $\vec{x} \equiv \mathbf{\Theta}_{\omega, k}^{-1} \vec{\mathbf{q}}_{1, \, k}$, $\vec{y} \equiv \mathbf{\Lambda}_k^H \, \vec{\mathbf{q}}_{1, \, k}$ and $a \equiv \beta_{\omega}$. Then, (\ref{demoduoneuser2}) is easily obtained.

\section*{Acknowledgment}
This work is supported by Vestel Electronics Inc., Manisa, 45030 Turkey.

\vspace{-1.5cm}
%


\begin{thebibliography}{1}

\bibitem{related0}
N. A. Johansson, et al., ``Radio Access for Ultra-reliable and Low-latency 5G Communications,''  \emph{IEEE International Conference on Communication Workshop (ICCW)}, pp.~1184--1189, 2015.
  
\bibitem{related4}
S. Bi, C. K. Ho, R. Zhang,, ``Wireless Powered Communication: Opportunities and Challenges,''  \emph{IEEE Communications Magazine}, vol.~53, no.~4, pp.~117--125, 2015.
 
\bibitem{burhan3}
B. Gulbahar, ``A Communication Theoretical Analysis of Multiple-access Channel Capacity in Magneto-inductive  Wireless Networks,'' \emph{submitted for publication}, 2016.
 
\bibitem{related2}
P. Grover, A. Sahai, ``Shannon meets Tesla: Wireless Information and Power Transfer,''  \emph{ISIT}, pp.~2363--2367, 2010.

\bibitem{related3}
L. R. Varshney, ``Transporting Information and Energy Simultaneously,''  \emph{IEEE Int. Symposium on Information Theory}, pp.~1612--1616, 2008.

\bibitem{related5}
R. Zhang, L. L. Yang, L. Hanzo, ``Energy Pattern Aided Simultaneous Wireless Information and Power Transfer,''  \emph{IEEE Journal on Selected Areas in Communications}, vol.~33, no.~8, pp.~1492--1504, 2015.

\bibitem{ref1}
B. Gulbahar and O. B. Akan, ``A Communication Theoretical Modeling and Analysis of Underwater Magneto-inductive Wireless Channels,'' \emph{IEEE Trans. on Wireless Communications}, vol.~11, no.~9, pp.~3326--3334,  2012.

\bibitem{ref2}
S. Kisseleff, I. F. Akyildiz and W. Gerstacker, ``Beamforming for Magnetic Induction based Wireless Power Transfer Systems with Multiple Receivers,'' \emph{arXiv preprint arXiv:1508.02566}, 2015.
 
\bibitem{ref4}
S. Li, Y. Sun and W. Shi, ``Capacity of Magnetic-Induction MIMO Communication for Wireless Underground Sensor Networks,'' \emph{International Journal of Distributed Sensor Networks}, Article ID 426324, 2015.
  
\bibitem{ref5}
B. Gulbahar, ``Theoretical Analysis of Magneto-Inductive THz Wireless Communications and Power Transfer with Multi-layer Graphene Nano-coils,'' to appear in \emph{IEEE Transactions on Molecular, Biological, and Multi-Scale Communications}, pp. 1--1, 2017.

\bibitem{related1}
L. Kisong, C. Dong-Ho, ``Simultaneous Information and Power Transfer Using Magnetic Resonance,''  \emph{ETRI Journal}, vol.~36, no.~5, pp.~808--818, 2014.

\bibitem{ref10}
H. Nguyen, J. I. Agbinya and J. Devlin, ``FPGA-Based Implementation of Multiple Modes in Near Field Inductive Communication Using Frequency Splitting and MIMO Configuration,''  \emph{IEEE Transactions on Circuits and Systems I}, vol.~62, no.~1, pp.~302--310, 2015.
 
\bibitem{ref12}
N. Ahmed et al., ``A Multi-coil Magneto-Inductive Transceiver for Low-cost Wireless Sensor Networks,''  \emph{IEEE Underwater Communications and Networking (UComms)}, pp.~1--5, 2014.
 
\bibitem{proakis}
J. G. Proakis, ``Digital communications,'' \emph{McGraw-Hill}, New York, 1995.
 
\bibitem{rosa}
E. B. Rosa and F. W. Grover, ``Formulas and Tables for the Calculation of Mutual and Self-Inductance,'' \emph{US Department of Commerce and Labor}, Bureau of Standards, 1912.
 
\bibitem{babic-validity}
S. Babic et al., ``Mutual Inductance Calculation Between Circular Filaments Arbitrarily Positioned in Space: Alternative to Grover's Formula,'' \emph{IEEE Transactions on Magnetics}, vol.~46, no.~9, pp.~3591-3600, 2010.

\bibitem{swiptrange}
{\color{black} 
I. Krikidis et al., ``Simultaneous Wireless Information and Power Transfer in Modern Communication Systems'', \emph{IEEE Communications Magazine}, vol. 5, no. 11, pp. 104--110, 2014.}

\end{thebibliography}
\end{document}